# Privacy-Preserving Medical Image Classification through Deep Learning and Matrix Decomposition

Andreea Bianca Popescu[1,2], Cosmin Ioan Nita[1], Ioana Antonia Taca[1,2], Anamaria Vizitiu[1,2], Lucian Mihai Itu[1,2]
[1]Transilvania University of Brasov, Brasov, Romania
[2]Siemens SRL, Brasov, Romania
andreeabianca.popescu@siemens.com, nita.cosmin.ioan@unitbv.ro, ioana_antonia29@yahoo.com, anamaria.vizitiu@siemens.com, lucian.itu@siemens.com

*Abstract*— Deep learning (DL)-based solutions have been extensively researched in the medical domain in recent years, enhancing the efficacy of diagnosis, planning, and treatment. Since the usage of health-related data is strictly regulated, processing medical records outside the hospital environment for developing and using DL models demands robust data protection measures. At the same time, it can be challenging to guarantee that a DL solution delivers a minimum level of performance when being trained on secured data, without being specifically designed for the given task. Our approach uses singular value decomposition (SVD) and principal component analysis (PCA) to obfuscate the medical images before employing them in the DL analysis. The capability of DL algorithms to extract relevant information from secured data is assessed on a task of angiographic view classification based on obfuscated frames. The security level is probed by simulated artificial intelligence (AI)-based reconstruction attacks, considering two threat actors with different prior knowledge of the targeted data. The degree of privacy is quantitatively measured using similarity indices. Although a trade-off between privacy and accuracy should be considered, the proposed technique allows for training the angiographic view classifier exclusively on secured data with satisfactory performance and with no computational overhead, model adaptation, or hyperparameter tuning. While the obfuscated medical image content is well protected against human perception, the hypothetical reconstruction attack proved that it is also difficult to recover the complete information of the original frames.

*Keywords—image obfuscation, singular value decomposition, principal component analysis, deep learning, medical imaging*

I. INTRODUCTION

In recent years, more and more deep learning-based solutions have been employed for solving various tasks in our daily life. The healthcare domain is no exception, as DL has achieved promising results in automating tedious tasks and facilitating doctors' work from diagnosis to treatment [1]. Due to the demand for such solutions, DLaaS (Deep Learning as a Service) platforms gained popularity. They emerged to assist those lacking the necessary expertise and processing resources to develop and deploy reliable DL models. Many hospitals and clinics are inclined nowadays to externalize specific tasks to DLaaS providers, both in terms of model training and new sample inference. However, the considerable amount of high-quality data typically required to develop and validate reliable solutions continues to be one of the biggest challenges. Compared to other domains where acquiring training data involves relatively low costs, patients' health-related information is a special class of personal data whose usage is strictly regulated by the General Data Protection Regulation (GDPR) and Health Insurance Portability and Accountability Act (HIPAA) in Europe and the United States, respectively. Data collection and outsourcing for DL purposes must adhere to privacy and personal information protection laws, necessitating appropriate anonymization. To address the data confidentiality concerns, there is a growing demand for effective privacy-preserving strategies that would enable DL model training based solely on anonymized, encrypted, or obfuscated data.

Fully homomorphic encryption (HE) is regarded as a straightforward method that enables a third party to process encrypted data without being aware of its content. Early attempts to incorporate HE into neural networks [2-4] showed promising achievements but also encountered roadblocks that limit their practical application. These obstacles included the constant requirement for communication between the model owner and the data provider, the impossibility of processing real numbers and the decreased prediction accuracy due to various approximations. Solutions to these problems have been proposed in numerous research papers, but none addressed the training phase of models on encrypted data. Due to the higher number of operations, and the longer runtime, the approaches focused on the inference phase. To overcome this shortcoming, we previously proposed a noise-free matrix-based homomorphic encryption alternative, which however has a lower security than conventional methods [5]. To mitigate the security vulnerability, we also integrated a numerical optimization method that facilitates training with a fixed number of operations, with an HE scheme, and an encoding technique that allows for computations on rational numbers [6]. However, a key disadvantage of this approach is the computational overhead. A recent survey [7] summarizes the various privacy-preserving primitives, such as multi-party computation, differential privacy, and federated learning, that could potentially be integrated into deep learning solutions, emphasizing that almost always a trade-off between security, performance and computational load should be considered.

Another approach focused on ensuring the confidentiality of image-type data consists of obfuscation. Various techniques have been proposed in past years, each with its own advantages and limitations. The most popular methods include mosaicking, blurring, and shuffling, which can obscure faces and numbers from human perception [8, 9]. Other innovative technique involves combining the pixels of two images or using a transformation generator [10, 11], the last being specifically designed for a medical application (MRI brain segmentation). The drawback of such methods is that they substantially impact the prediction performance. Some were also proven vulnerable to reconstruction attacks. Generative adversarial networks are typically the foundation of the techniques that aim to produce visually pleasing images [12, 13]. These models can be used to generate face detection data while offering protection against face recognition. Our previous work on image obfuscation started with analyzing the effect of bijective and non-bijective functions, when pixel intensity shuffling is used to hide the content of an image [14]. The study proved that non-bijectivity increased security without significantly impacting classification performance. Then, we combined the encoding of a variational autoencoder with the non-bijective shuffling to increase both the accuracy and the privacy [15].

In this paper, we propose image obfuscation approaches based on singular value decomposition and on principal component analysis, respectively. This research aims to determine, for each of the proposed techniques, the degree to which they meet the following requirements: (i) obfuscate sensitive image content from human perception, (ii) hinder AI-based image recovery, and (iii) facilitate the usage of obfuscated images in DL model training. The same methodology and experimental setup as in our previous work are used to enable a fair comparison. The experiments are constructed to reflect the perspective of a clinical user that relies on an external party (i.e., the DLaaS provider) to develop a model that solves a basic but tedious task (in this case, angiographic view classification). The DLaaS provider can access, for this purpose, only obfuscated training data. The model is then made available to the hospital in the cloud, where new images will be sent for inference in their obfuscated state. On the other hand, a threat actor (e.g., another healthcare company, the DLaaS provider, or an interceptor) wants to access the non-obfuscated version of the patients' data. The threat actor conducts an AI-based reconstruction attack to recover the image's original state. Two different levels are assumed for the attacker's knowledge: the reconstruction attempts are performed with a model trained on generic medical images, and with a model trained on angiographies similar to the classification dataset (i.e., the data targeted by the threat actor).

The rest of the manuscript is organized as follows. The obfuscation approaches and the methods and materials used to analyze their privacy-preserving properties in the context of DL-based solution development are presented in Section II. Section III describes the experiments performed from the utility and privacy perspectives, along with the findings. In Section IV, we compare the results with our previous work, propose subsequent research directions considering the limitations of the current experimental setup, and draw conclusions.

## II. METHODS AND MATERIALS

### A. Image Obfuscation based on Singular Value Decomposition

Singular value decomposition (SVD) is a method for matrix factorization that generalizes eigendecomposition to matrices of arbitrary dimensions. Applying this factorization for image obfuscation consists of treating the entire image as a matrix and using the decomposition formula (1), where $I$ is the image matrix, $U$ is a square matrix containing the eigenvectors of $I \cdot I^T$, $V^H$ is the Hermitian transpose of $V$, including the eigenvectors of $I^T \cdot I$ and $S$ is a diagonal matrix that contains the singular values of $I$, uniquely determined by the image $I$. $I^T$ denotes the transpose of the image matrix.

$$I = U \cdot S \cdot V^H \quad (1)$$

The intuition on which the obfuscation approach is based on is that discarding the $S$ matrix would ensure image privacy by removing the underlying information up to a certain degree. At the same time, features of the original image are stored in $U$ and $V^H$ enabling the successful development of DL-based solutions. Thus, the scenarios where the obfuscated image, $I^O$, is computed as $I^O = U$, $I^O = V^H$, and $I^O = U+V^H$ are analyzed in terms of utility, privacy level and protection against AI-based reconstruction. A comparison between an angiographic frame and the obfuscated counterparts for the three SVD-based approaches is depicted in Fig. 1.

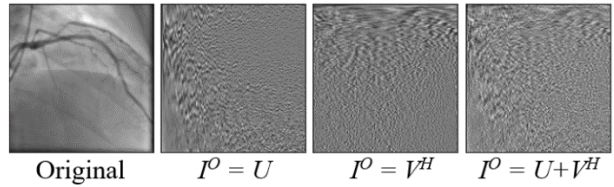

Fig. 1. Comparison between an original angiographic frame and different SVD-based obfuscations.

### B. Image Obfuscation based on Principal Component Analysis

The principal component analysis is a technique based on the singular value decomposition of a matrix. It is typically used to examine a large dataset, with a high number of features per sample, to reduce the dimensionality while conserving as much information as possible. To make use of this technique's properties for image obfuscation, the following steps are applied:

1. The first $N$ principal components are extracted based on the training set.
2. The previously identified components are extracted from each image of the dataset.
3. Each image's obfuscated counterpart is computed by reconstructing the frame from the extracted components, employing the inverse transformation.

The intuition is that training a DL model with this type of data would lead to higher performance because more of the original information is stored in the obfuscated frames, as the matrix now consists of actual pixel intensities and not eigenvectors. However, it is expected that this will negatively impact the privacy level, hence an additional operation is inserted between steps 2 and 3: randomly shuffling the principal components. This approach will be abbreviated PCA-SC (shuffled components).

Nevertheless, since data from the same distribution is required to successfully train the DL model, the same principal components and shuffling must be applied to all the images in the dataset. Examples of images obfuscated with the PCA-based techniques for different numbers of principal components are shown in Fig. 2.

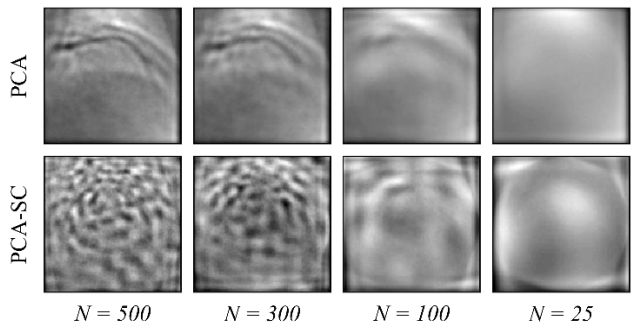

Fig. 2. Comparison between different PCA-based obfuscations.

A critical remark is that both SVD-based and PCA-based approaches are deterministic, meaning that a frame will always be obfuscated identically. This makes them vulnerable to re-identification attacks where the threat actor, who has a pair of original-obfuscated frames, wants to know if that particular image can be found in a dataset. Hence, the threat actor may compare all images against the available obfuscated sample. The solution would be to add some randomness during obfuscation, which will always lead to a different result. To achieve this behavior, one-third of the original image rows are randomly selected, and one pixel is replaced on each row. The pixel's position and its new value are also randomly chosen for each image.

*C. Utility and Privacy Evaluations*

The utility of the images obfuscated with the approaches described above must be assessed since these techniques lead to the partial loss of content from the original sample. The evaluation is carried out by training the same DL architecture with samples that have been altered using each of the methods discussed above and with the original data separately. The presumed scenario is that of a hospital planning to automate the X-ray coronary angiography view classification. We employ a collection of frames from an internal dataset that displays either the right coronary artery (RCA) or the left coronary artery (LCA). The dataset totals 3280 coronary angiographies resized to 128 x 128 resolution and has an equal number of samples for each class. The min-max scaling technique is used to normalize the pixel intensities in the [0, 1] interval. The training phase is based on 1980 angiographic frames, augmented through shifting, flipping, zooming and rotation. A subset of 600 images is used for validation, while a separate subset of 700 test images is kept for evaluation. A network composed of convolutional and fully connected layers is used to solve the task. A particularity of the architecture is the use of local normalization layers [16] to increase the robustness of the model to image alteration. More details regarding the dataset, the architecture and the training process are described in [15].

To assess the security against human perception, the structural similarity metric (SSIM) and peak signal-to-noise ratio (PSNR) between the original and the obfuscated angiographies of the testing subset are computed. SSIM can range from 0 to 1, with 0 denoting no structural similarity and 1 denoting identical frames. According to [10], it can quantify image privacy; thus, lower values will be interpreted as increased security. The PSNR is expressed in decibels, good quality images (with an 8-bit bit depth) often having values between 30 and 50 dB. Values below the lower threshold signify that the image's content is successfully masked.

Two scenarios are simulated to measure the security against AI-based reconstruction attacks. The following steps describe the method that the threat actor is adopting to obtain access to the data sent by the hospital: (i) apply the same obfuscation technique as the hospital on his own dataset; (ii) train a DL model to recover the original images from the obfuscated counterparts; (iii) use the model to reconstruct the original content of the intercepted frames originating from the hospital. Firstly, it is assumed that the threat actor is aware that the targeted data consists of medical image acquisitions but does not know that they are angiographic frames ($S_1$). In this case, the reconstruction model is trained using a generic dataset containing six different classes of medical images (abdomen CT, breast MRI, CXR, chest CT, hand radiography, head CT). The more vulnerable scenario is the one where the threat actor trains the network using his own dataset of coronary angiographies, i.e., the actor is aware that the target data consists of such images ($S_2$). Regardless of the scenario, the images are resized to 128 x 128 pixels and normalized to the [0, 1] range before being utilized in training. The test subset of the hospital data is also employed here to evaluate the reconstruction. SSIM and PSNR are computed between the original and reconstructed frames for a quantifiable evaluation. A U-net [17] architecture is employed for reconstruction. The model and the datasets are described in detail in [15].

III. EXPERIMENTS AND RESULTS

*A. Benchmark*

Before training the classifier with obfuscated images, the model is evaluated on the original data. The accuracy on the test subset, also reported in [15], is 97.57%. To measure the information loss due to random pixels' replacement as a solution for determinism, an experiment with the original data where one-third of the rows had one pixel randomly changed was run. This model's accuracy was 96.71%. Hence, this step does not significantly impact the performance as the less than 1% difference could be considered originating also from other random processes during the training (random initialization). We consider this value to be the comparison benchmark.

*B. Utility and Privacy Assessment of the Approach based on Singular Value Decomposition*

As described above, three possible obfuscation approaches are derived from the singular value decomposition. The classification results for each of them are shown in Table I.

TABLE I.   CLASSIFICATION PERFORMANCE FOR SVD OBFUSCATION

| Obfuscation | $I^O = U$ | $I^O = V^H$ | $I^O = U+V^H$ |
|---|---|---|---|
| Accuracy [%] | 87.00 | 85.71 | 91.85 |

The $U$ and $V^H$ components preserve a similar amount of information, and the models trained with this type of data achieve satisfactory results. They are, however, outperformed by the model trained with images that are obfuscated by summing the two matrices, for which the accuracy dropped by less than 5% compared to the benchmark. The corresponding similarity metrics are shown in the first two rows of Table II. As expected, the numbers are higher when both matrices are used, but they still indicate that it is impossible to distinguish the image's content through visual inspection. The reconstruction results are also synthesized in Table II.

TABLE II. SIMILARITY METRICS FOR SVD OBFUSCATION

| Obfuscation | $I^O = U$ | $I^O = V^H$ | $I^O = U+V^H$ |
|---|---|---|---|
| SSIM Obfuscation | 0.2767 | 0.2740 | 0.3155 |
| PSNR Obfuscation [dB] | 19.89 | 19.88 | 20.13 |
| SSIM Reconstruction $S_1$ | 0.6930 | 0.6850 | 0.6935 |
| PSNR Reconstruction $S_1$ [dB] | 18.99 | 18.45 | 19.31 |
| SSIM Reconstruction $S_2$ | 0.7138 | 0.7126 | 0.7121 |
| PSNR Reconstruction $S_2$ [dB] | 22.00 | 22.11 | 21.46 |

The SSIM is higher than when computed for the obfuscated samples, indicating that some structural information is recovered through reconstruction. As expected, it is higher for S2, where the data used to train the reconstructed model stems from a distribution similar to that of the inferred samples. However, the PSNR remains low, confirming that the content of the reconstructed images is indistinguishable. The U-net architecture cannot map the eigenvectors to the actual pixel intensities of the original image in either of the scenarios. Fig. 3 depicts the images reconstructed in the $S_2$ scenario. Those corresponding to $S_1$ looked very similar and thus were omitted.

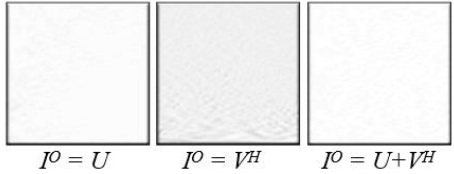

Fig. 3. Reconstruction attempts for different SVD-based obfuscations.

### C. Utility and Privacy Assessment of the approach based on Principal Component Analysis

The PCA-based methods allow the user to adjust the privacy-accuracy trade-off. By selecting an appropriate number of principal components, *N*, one can opt for a higher level of security while ensuring a satisfactory performance. The maximum number of components is represented by the total number of pixels, which for a 128 x 128 image is 16,384. We performed multiple experiments reducing the number of components from 500 to 100 by 50 and from 100 to 25 by 25. The classification results for the PCA and the PCA-SC approaches are compared in Fig. 4.

The highest accuracy is 94.71%, only 2% lower than the benchmark, but it is achieved for the PCA approach using 500 components, which only partially blurs the vessels in the angiography. A compromise value of *N=100* can be selected, considering that the accuracy is still greater than 90% while the privacy level is satisfactory. For the PCA-SC technique, the best performance was 89.57% accuracy when 450 components were used in obfuscation. This case also led to a high privacy level, with the vessels in the image being completely hidden. It can be noticed that fewer components do not automatically imply a lower accuracy. However, a descending trend can be noticed when the obfuscated image is obtained directly from the extracted components. This tendency is less evident when the components are shuffled before using the inverse transformation. This behavior is expected because fewer components reduce entropy while providing less relevant information.

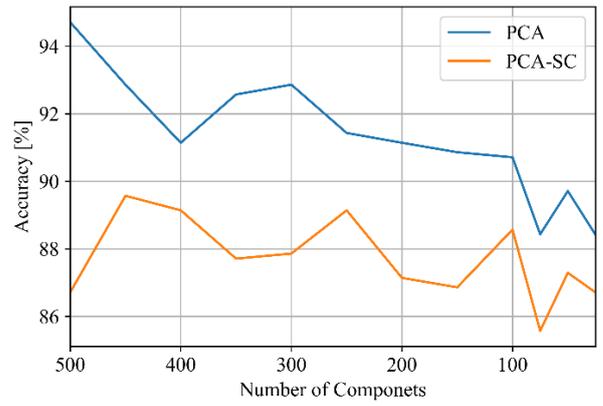

Fig. 4. Comparison between classification performance achieved when PCA and PCA-SC techniques are employed for image obfuscation.

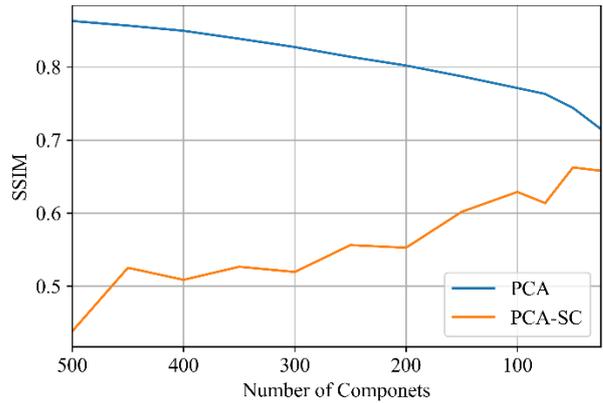

Fig. 5. Comparison between PCA and PCA-SC obfuscation techniques with respect to the SSIM between original and obfuscated angiographies.

Thus, PCA-SC is more security-oriented than PCA, which is more focused on the performance. For the PCA-based approach, the structural similarity and the image quality decrease as the number of components decreases (Fig. 5, Fig. 6). On the contrary, the direct proportionality between the number of components and the randomness in the PCA-SC scenario can explain the similarity's increase for fewer components. Although the SSIM is more sensitive to any structural change, the PSNR provides an overall quantification for the image quality, which, for *N* under 250, is low regardless of the method.

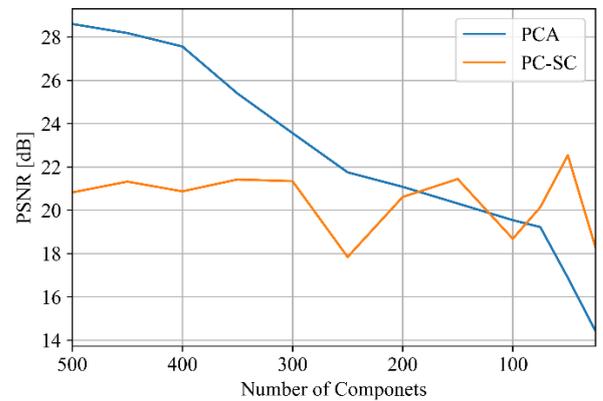

Fig. 6. Comparison between PCA and PCA-SC obfuscation techniques with respect to the PSNR between original and obfuscated angiographies.

The same reconstruction attack scenarios were simulated, Fig. 7 depicting samples of recovered frames, starting from different levels of obfuscation. While the model can improve to some extent the image quality, starting from the details that are still present in the angiographies obfuscated with PCA, the PCA-SC protects the model against this attack.

The privacy metrics shown in Fig. 8 and Fig. 9 corroborate the idea that the threat actor's awareness of the content of the targeted images can increase the likelihood of successful data recovery since higher values were registered for nearly all experiments under scenario $S_2$.

IV. DISCUSSION AND CONCLUSIONS

Since it is challenging to compare the outcomes of this research with other similar works in the literature because of the differences in the datasets and methodologies employed, we briefly compare the results with what we achieved in our prior work [14,15].

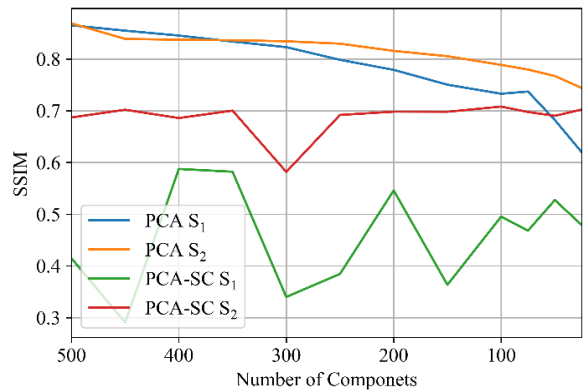

Fig. 8. Comparison between PCA and PCA-SC obfuscation techniques with respect to the SSIM between original and reconstructed angiographies in two attack scenarios.

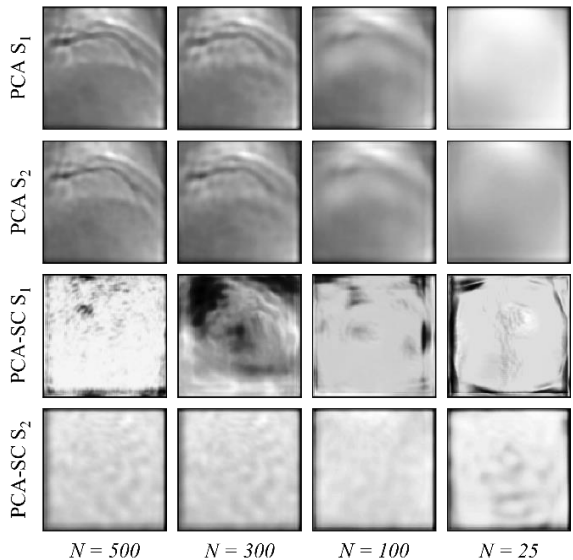

Fig. 7. Reconstruction attempts for different PCA-based obfuscations in two attack scenarios.

Previously, the best classification accuracy obtained for obfuscated images on this dataset was 93.71% when the angiographies were altered using a variational autoencoder. However, to some extent, the vessels were distinguishable in the obfuscated frame and could also be recovered, when employing the same reconstruction architecture as the one in the current experiments. The highest accuracy on the frames obfuscated with the methods described above was 94.71% (PCA, 500 components), but this is again the case of a low level of alteration, where many details are still visible in the image. Nevertheless, the highest performance achieved in this work when the image's content is entirely undistinguishable was 91.85% (SVD, $I^O = U+V^H$) which is almost 10% higher than the accuracy obtained with the secure algorithm that we previously proposed (that combined VAE with non-bijective pixel intensities shuffling). The overall privacy and reconstruction metrics were lower in the current work, indicating that the methods based on SVD and PCA provide better security than previously analyzed approaches.

It is important to note that the purpose of the experiments is not to achieve state-of-the-art performance in angiographic view classification on obfuscated data, which is why the same architecture was employed for all types of input data. Future research should explore how much performance can be improved by adjusting the network specifically for obfuscated images, and what deep learning techniques are the most suitable for building models using input undistinguishable to humans. From our experiments, for the SVD approach, using in conjunction the $U$ and $V^H$ component, either in a two inputs model where parallel networks analyze the components, or as a two-channel input in a one-branch network, increases the performance by ~2%, when compared to the setting where $U$ and $V^H$ are summed.

On the other hand, the fact that the reconstruction failed in most of the cases does not imply that the method is perfectly secure against any reconstruction attack but just that the employed architecture (that previously managed to partially decode images obfuscated with other methods) cannot recover this specific dataset when it is obfuscated using the methods presented in this manuscript. As a future step, an analysis across multiple datasets and multiple attack configurations (threat actor's knowledge level, attack techniques, architectures employed) would be beneficial for assessing the true strength of these obfuscation techniques against AI-based reconstruction attempts.

Although some of the accuracy scores are 10% below the benchmark, as the dataset is balanced and the two classes are predicted equally, the performance can still be considered satisfactory. Even though data security was ensured during most experiments, the few cases where the classification results were very close to the benchmark with the cost of poor obfuscation emphasized the need for careful consideration of the privacy-accuracy trade-off. Thus, the analyzed techniques keep image content confidential and protect it against human and AI perception, while ensuring its utility in developing deep learning solutions.

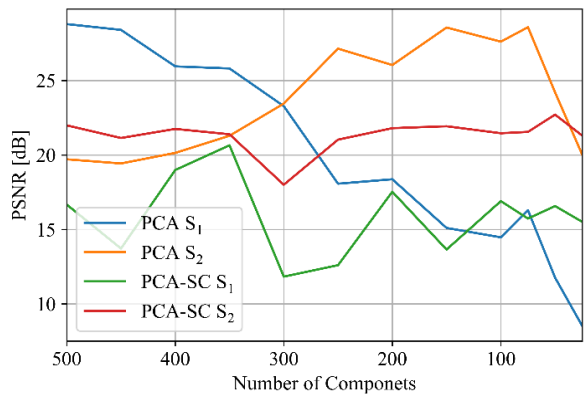

Fig. 9. Comparison between PCA and PCA-SC obfuscation techniques with respect to the PSNR between original and reconstructed angiographies in two attack scenarios.

ACKNOWLEDGMENT

This work was supported by a grant of the Romanian National Authority for Scientific Research and Innovation, CCCDI – UEFISCDI, project number ERANET-PERMED-RESPECT, within PNCDI III. The research leading to these results has received funding from the EEA Grants 592 2014-2021, under Project contract no. 33/2021.

REFERENCES

[1] R. Miotto, F. Wang, S. Wang, X. Jiang, and J.T. Dudley, "Deep learning for healthcare: review, opportunities and challenges", in Briefings in bioinformatics, vol. 19 (6), pp. 1236-1246, 2018.

[2] C. Orlandi, A. Piva, and M. Barni, "Oblivious neural network computing via homomorphic encryption", EURASIP Journal on Information Security, pp. 1-11, 2007.

[3] R. Gilad-Bachrach, N. Dowlin, K. Laine, K. Lauter, M. Naehrig, and J. Wernsing, "Cryptonets: Applying neural networks to encrypted data with high throughput and accuracy", in International conference on machine learning, pp. 201-210, 2016.

[4] E. Hesamifard, H. Takabi, and M. Ghasemi, "Cryptodl: Deep neural networks over encrypted data", arXiv preprint arXiv:1711.05189, 2017.

[5] A. Vizitiu, C.I. Niţă, A. Puiu, C. Suciu, L.M. Itu, "Applying Deep Neural Networks over Homomorphic Encrypted Medical Data", in Computational and mathematical methods in medicine, 2020.

[6] A.B. Popescu, I.A. Taca, C.I. Nita, A. Vizitiu, R. Demeter, C. Suciu, et al., "Privacy preserving classification of eeg data using machine learning and homomorphic encryption", in Applied Sciences, 2021.

[7] H. Chen, S.U. Hussain, F. Boemer, E. Stapf, A.R. Sadeghi, et al., "Developing privacy-preserving AI systems: The lessons learned", in 57th ACM/IEEE Design Automation Conference (DAC). IEEE, 2020.

[8] R. McPherson, R. Shokri, and V. Shmatikov, "Defeating image obfuscation with deep learning", arXiv preprint arXiv:1609.00408, 2016.

[9] T. Zhang, Z. He, R. B. Lee, "Privacy-preserving machine learning through data obfuscation", arXiv preprint arXiv:1807.01860, 2018.

[10] M. Raynal, R. Achanta, and M. Humbert, "Image Obfuscation for Privacy-Preserving Machine Learning", arXiv preprint arXiv:2010.10139, 2020.

[11] B.N. Kim, J. Dolz, C. Desrosiers, and P. M. Jodoin, "Privacy Preserving for Medical Image Analysis via Non-Linear Deformation Proxy", arXiv preprint arXiv:2011.12835

[12] T. Li and M. S. Choi, "DeepBlur: A simple and effective method for natural image obfuscation", arXiv preprint arXiv:2104.02655 1, 2021.

[13] J.W. Chen, L.J. Chen, C.M. Yu, and C.S. Lu, "Perceptual Indistinguishability-Net (PI-Net): Facial Image Obfuscation with Manipulable Semantics", in Proceedings of the IEEE/CVF Conference on Computer Vision and Pattern Recognition, pp. 6478-6487, 2021.

[14] A.B. Popescu, C.I. Nita, I.A. Taca, A. Vizitiu and L.M. Itu, "Non-bijectivity-based image obfuscation method for deep learning based medical applications", 2022 International Conference on Development and Application Systems (DAS), Suceava, Romania, pp. 132-136, 2022.

[15] A.B. Popescu, I.A. Taca, A. Vizitiu, C.I. Nita, C. Suciu, et al., "Obfuscation Algorithm for Privacy-Preserving Deep Learning-Based Medical Image Analysis", in Applied Sciences, 2022.

[16] B. Yin, H.S. Scholte, and S. Bohté, "LocalNorm: Robust Image Classification Through Dynamically Regularized Normalization", in Artificial Neural Networks and Machine Learning–ICANN, Bratislava, Slovakia, Proceedings, Springer International Publishing, pp. 240-252, 2021.

[17] O. Ronneberger, P. Fischer and T. Brox, "U-net: Convolutional networks for biomedical image segmentation", in Medical Image Computing and Computer-Assisted Intervention–MICCAI 2015: 18th International Conference, Munich, Germany, pp. 234-241, 2015.